\documentstyle[times,psfig,draft]{basi}
\def\bi{\bibitem{}}

\def\ni{\noindent}
\def\beb{\begin{thebibliography}{10}}
\def\eeb{\end{thebibliography}}
\def\bei{\begin{itemize}}
\def\eei{\end{itemize}}
\def\bef{\begin{figure}}
\def\eef{\end{figure}}
\def\ben{\begin{enumerate}}
\def\een{\end{enumerate}}
\def\beq{\begin{equation}}
\def\eeq{\end{equation}}
\def\ber{\begin{eqnarray}}
\def\eer{\end{eqnarray}}
\def\edo{\end{document}}      

\def\Bb{{\bf B}}
\def\pa{\partial}
\def\vb{{\bf v}}

\newcommand{\gcc}{{\rm g} \, {\rm cm}^{-3}}

\begin{document}
\title[magnetar fields]{The {\it Magnetar} Fields}
\author[Sushan Konar]%
       {Sushan Konar\thanks{e-mail:sushan@cts.iitkgp.ernet.in} \\ 
        i. Department of Physics \& Meteorology, 
        ii. Centre for Theoretical Studies, \\
        Indian Institute of Technology, Kharagpur 721302}
%\pubyear{2001}
%\volume{29}
%\setcounter{page}{17}
%\date{Received 2001 May 30; accepted 2001 June 07}
\maketitle
\label{firstpage}
\begin{abstract}
We discuss the nature of evolution of the magnetic field in {\it magnetars}.
\end{abstract}


\ni There is growing evidence that soft gamma-ray repeaters (SGR) and 
anomalous X-ray pulsars (AXP) are isolated neutron stars with superstore 
magnetic fields ($B > 10^{14}$~G), commonly known as {\it magnetars}. 
Recently, the detection of a 5 Kiev absorption feature from SGR 1806-20 has 
been identified as proton cyclotron resonance, offering a direct measurement 
of the magnetic field to be $\sim 10^{15}$~G (Ibrahim, Swank \&  Parke 2003).
Decay of this strong magnetic field is supposed to power these {\it magnetars}.
The strong fields affect the physical processes in the stellar interior and 
consequently influence the evolution of the magnetic field itself. 

\ni 
The ratio between the Landau level spacing and the Fermi energy for the 
electrons is,
\beq
\frac{E_F}{\Delta E_{\rm LL}}
= 2.5 \, \mu^{-1/3} \rho_{15}^{1/3} {\cal B}_{15}^{-1}, \,\; \;\;\;
(\mbox{$\rho = \rho_{15} 10^{15}\gcc$, ${\cal B} = {\cal B}_{15} 10^{15}$~G})\,.
\eeq
As seen from Fig.1A, this ratio is smaller than unity for a large range of 
densities where the system goes into the quantum Hall regime. For neutron 
stars born with strong fields the electron fluid is already in this regime 
at birth. Then the outer regions of the star may not form a crystalline 
solid upon cooling, greatly affecting the transport properties. 

\bef
\begin{center}{
\psfig{file=fig01.ps,width=150pt,angle=-90}
\vspace{-3.95cm}
\hspace{4.50cm}
\psfig{file=fig02.ps,width=150pt,angle=-90}
}\end{center}
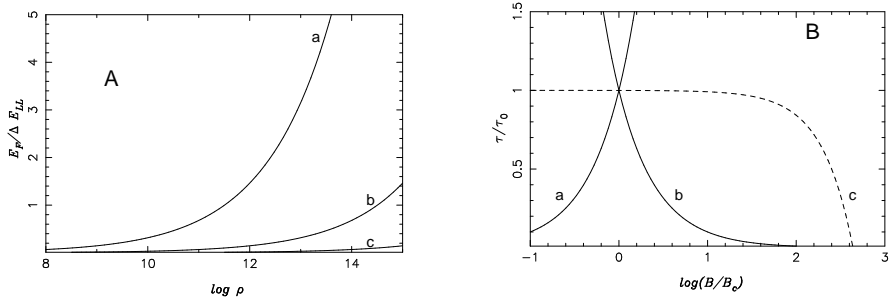
\caption[]{{\bf (A)} - Ratio of Fermi Energy to Landau level spacing for 
electrons. Curves $a,b,c$ correspond to $B = 10^{16}, 10^{15}, 10^{14}$~G. 
{\bf (B)} - Ratio of strong-field ohmic dissipation time-scale to that for 
zero magnetic field. Curves $a,b$ correspond to conductivities taken from 
Yakovlev (1984) and Ghost et al. (2002). Curve $c$ shows the expected behaviour.}
\label{fig01}
\eef

\ni As the crustal lattice yields under the pressure exerted by a $\sim 10^{15}$~G 
field, the currents need to be anchored in the deeper interior of the star, 
comprising of an n-p-e plasma. Assuming the plasma to be non-superfluid the 
evolution of the magnetic field would be governed by the induction equation 
(Goldreich \& Reisenegger, 1992)
\beq
\frac{\pa \Bb}{\pa t} 
= - c \nabla \times \left( \frac{\bf J}{\sigma_0} \right)
  + \nabla \times (\vb \times \Bb) 
  - \frac{m_p/\tau_{pn} - m_e/\tau_{en}}{m_p/\tau_{pn} + m_e/\tau_{en}} 
    \nabla \times \left(\frac{{\bf J} \times \Bb}{n_ce} \right) \,,
\eeq
where $\sigma_0$ is the homogeneous electrical conductivity of the medium in
absence of a magnetic field. Since, the magnetar fields are strongly quantizing 
the transport properties would be highly anisotropic as well as field-dependent 
(Yakovlev, 1984; Ghosh et.al, 2002). This renders the ohmic dissipation time-scale 
of the internal currents field dependent as shown in Fig.1B. However, we expect the 
field to decay faster for stronger fields and the dependence of $\tau_{\rm ohmic}$ 
should be more like the dashed curve in Fig.1B (obtained by using 
$\sigma (B) = \sigma_0 (1 - \beta (B/B_c)^\alpha)$ where $B_c$ is the electron
critical field) and contrary to that obtained using the electrical conductivity 
of either Yakovlev (1984) or Ghosh (2002). 

\ni
Evidently, the terms corresponding to Hall drift and Ambipolar diffusion would
be important for the magnetar field evolution. The anisotropy in the conductivity
would give rise to strong Hall currents inducing rapid transfer of energy from 
dipolar component of the field to higher multi-poles. Since, currents of higher 
order multi-poles dissipate faster (Mitra, Konar \& Bhattacharya 1999) the field 
would decay rapidly because of this. The details of this work can be found in Konar 
(2003).

\beb
\bi Ghosh S. et al., 2002, \newblock {\it IJMPD}, {\bf 11}, 843
\bi Goldreich P., Reisenegger A., 1992, \newblock {\it ApJ}, {\bf 395}, 250
\bi Ibrahim A.~I., Swank J.~H., Parke W., 2003, \newblock {\it ApJL}, {\bf 584},17
\bi Konar, S., 2003, \newblock {\it in preparation}
\bi Mitra D.,Konar S., Bhattacharya D., 1999, \newblock {\it MNRAS}, {\bf 307}, 459
\bi Yakovlev D.~G., 1984, \newblock {\it ApSS}, {\bf 98}, 37
\eeb
\label{lastpage}

\end{document}